\documentclass[prd,twocolumn,showpacs,floatfix,amsmath,nofootinbib,amssymb,floatfix]{revtex4}
\usepackage{graphicx,color,dcolumn,booktabs,bm}
\usepackage{longtable,lscape}
\usepackage{txfonts}
\usepackage{overpic}
\usepackage{amssymb}
\usepackage{indentfirst}
\usepackage{feynmf}   
\usepackage{slashed}  
\usepackage{cases}
\usepackage{color}
\usepackage{multirow}
\usepackage{epstopdf}
\usepackage{xcolor}

\usepackage{graphicx,color,dcolumn,booktabs,bm}
\usepackage[colorlinks,
            citecolor=blue,
            anchorcolor=red,
            menucolor=red,
            linkcolor=red,
            filecolor=red,
            runcolor=red,
            urlcolor=blue,
            frenchlinks=red]{hyperref}

\begin{document}
\title{Arrangement for $K_2^*$ meson family}
\author{Ting-Yan Li$^{1,2}$}\email{litingyan1213@163.com}\author{Ya-rong Wang$^{1,2}$}\email{nanoshine@foxmail.com}
\author{Cheng-qun Pang$^{1,2,3}$\footnote{Corresponding author}}\email{pcq@qhnu.edu.cn}
\affiliation{$^1$College of Physics and Electronic Information Engineering, Qinghai Normal University, Xining 810000, China\\$^2$Joint Research Center for Physics,
Lanzhou University and Qinghai Normal University,
Xining 810000, China \\$^3$Lanzhou Center for Theoretical Physics, Key Laboratory of Theoretical Physics of Gansu Province, Lanzhou University, Lanzhou, Gansu 730000, China}
\vspace{2cm}
\begin{abstract}
Two observed  structures with $M=1868 \pm 8^{+ 40}_{- 57}$ MeV and $M=2073 \pm 94^{+ 245}_{- 240}$  MeV are the same states ($K_2^*(1980)$) in PDG.The analysis of the mass spectrum and the calculation of the strong decay of $K_2^*$ mesons support the low mass state of $K_2^*(1980)$ as $2^3P_2$ and the high mass state of $K_2^*(1980)$ as $1^3F_2$ in this letter. This analysis brings us very important criterion for the assignment of the observed $K_2^* (1980)$ and experimental findings for this assignment is suggested. Additionally, prediction of some partial decay widths are made on the high excitations of $K_2^*$ family. This study is crucial to establishing and searching for their higher excitations in the future.

\end{abstract}

\pacs{13.25.Es, 14.40.Df}
\maketitle

\section{Introduction}

$K_2^*$ meson family is a crucial component of kaon family. There are two members in this $K_2^*$ mesons: $K_2^*(1430)$ and $K_2^* (1980)$. $K_2^*(1430)$ is well established as the ground state of $K_2^*$ meson family as a $1^3P_2$ assignment.
$ K_2^* (1980)$ is now listed in PDG \cite{Zyla:2020zbs} with an average mass and width of
$1995^{+60}_{-50}$ MeV and $349^{+50}_{-30}$ MeV.
If $K_2^* (1980)$ is a $2^3P_2$ state or a $1^3F_2$ state, this arouses our attention.

 $K_2^* (1980)$ meson was reported by LASS in 1987 and 1989 in $K^-p\to\bar{K}^0\pi^+\pi^-n$ and $K^-p\to\bar{K}^0\pi^-p$  processes, which mass is $1973 \pm 8\pm 25$ MeV and corresponding width is $373 \pm 33\pm 60$ MeV(in 1989, LASS gave the resonance parameters  of $K_2^* (1980)$ with $\Gamma=1978\pm40$ MeV and $M=398\pm47$ MeV, respectively) \cite{ASTON1987693,Zyla:2020zbs}. It is likely to be the candidate 
of $2^3P_2$ state or $1^3F_2$ state.

 Recently, the BESIII Collaboration observed $K_2^* (1980)$ in the $K\pi$ channel in the process $J/\psi\to K^+K^-\pi^0$.They obtained two solutions when fitting the experimental data, $M=1817\pm 11$ MeV and $\Gamma=312\pm28$ MeV, or $M=1868\pm 8^{+40}_{-51}$ MeV and $\Gamma=272\pm 24^{+50}_{-15}$ MeV \cite{BESIII:2019apb}. Its mass is around 250 MeV lower than the $2073 \pm 94^{+ 245}_{- 240}$  MeV that detected by the LHCb Collaboration \cite{LHCb:2016nsl}. The resonances of $K_2^*(1980)$ were later provided by LHCb Collaboration. They regarded $K_2^*(1980)$ as $2^3P_2$ state which $J^P = 2^+$, the mass is of $1988 \pm {22^{+ 194}_ {- 31}}$ MeV and the width is of $318 \pm {82 ^{+ 481}_{- 101}}$ MeV \cite{LHCb:2021uow}.
Recent observations of $K_2^* (1980)$ by the BESIII Collaboration by means of  
partial wave analysis of $\psi(3686)\to K^{+}K^{-}\eta$ also gave the resonance parameters $M=2046^{+17+67}_{-16-15}$ MeV and $\Gamma=408^{+38+72}_{-34-44}$ MeV \cite{BESIII:2019dme}.
 Are these structures the same state?
 
 \begin{figure}
\includegraphics[width=6.3cm]{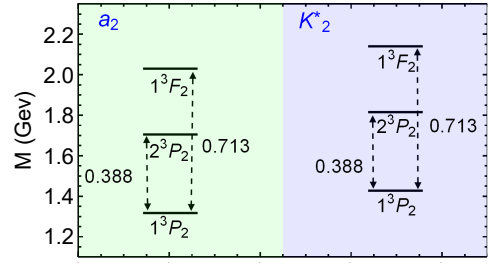}
\caption{The mass gap comparison between $K_2^*$ and $a_2$ family. The left part of the figure is the mass gap of $a_2$ family and the right part of the figure is the mass gap of $K_2^*$ family.}
\label{mass}
\end{figure}

We contrast the $a_2$ and $K_2$ families in Fig. \ref{mass} to find the solution for this query. In the $a_2$ family, The mass difference between $1^3P_2$ and $2^3P_2$ states is $388$ MeV. Additionally, there is a mass gap of $713$ MeV  between $1^3P_2(a_2(1320))$  and $1^3F_2(a_2(2030))$ states. If we put this mass gap into $K_2$ family, and use $1427.3$ MeV as the mass of  $1^3P_2(K^*_2(1430))$ state, then the mass of $2^3P_2$ state will be $1815.3$ MeV, which is extremely close to the experimental value $M=1817\pm 11$ MeV(or solution two $1868 \pm 8^{+ 40}_{- 57}$ MeV) \cite{Zyla:2020zbs}. The mass of $1^3F_2$ state will be $2140.3$ MeV just consistent with $2073 \pm 94^{+ 245}_{- 240}$  MeV \cite{Zyla:2020zbs}. Following this analysis, the structure with $M=1817\pm 11$ MeV(or second solution $1868 \pm 8^{+ 40}_{- 57}$ MeV, in this work, we name it $K_2^*(1870)$) \cite{Zyla:2020zbs}  should be $2^3P_2$ state and the state with $M=2073 \pm 94^{+ 245}_{- 240}$  MeV \cite{Zyla:2020zbs} (in this work, we name it $K_2^*(2070)$) could be $K_2^*(1^3F_2)$ state.

 For testing the proposal of assignments of $K_2^*(1870)$ and $K_2^*(2070)$, in the following, we will give the mass and decay width of $K_2^*(2^3P_2)$ and $K_2^*(1^3F_2)$ via the mass spectrum and two body strong decay of $K_2^*$. In the same time, the property of the higher excited $K_2^*$ states will be investigated.

Godfrey and Isgur raised the GI model for describing relativistic meson spectra in 1985 \cite{Godfrey:1985xj}. Thirty years later, Song $et~al.$ developed the modified GI (MGI) model taking into account  the screening effect in GI model \cite{Song:2015nia,Song:2015fha}.  This modified GI model were used to calculate the meson spectra of high excited states \cite{Pang:2018gcn,Wang:2018rjg,Pang:2019ttv,Wang:2019mhs,Li:2021qgz}. In this work, we calculate the mass spectra of the $K_2^*$ family by applying the modified GI model \cite{Song:2015nia,Song:2015fha}. The spatial wave functions obtained by the modified GI model can be  taken as input when we studying $K_2^*$ family's Okubo-Zweig-Iizuka (OZI)-allowed two-body strong decays adopting the quark pair creation (QPC) model, which was proposed in Ref. \cite{Micu:1968mk} and extensively applied to studies of other hadrons in Refs. \cite{LeYaouanc:1972ae,
vanBeveren:1979bd,vanBeveren:1982qb,LeYaouanc:1988fx,roberts,Capstick:1993kb,Blundell:1995ev,Ackleh:1996yt,Capstick:1996ib,Bonnaz:2001aj,
Close:2005se,Zhang:2006yj,Lu:2006ry,Sun:2009tg,Liu:2009fe,Sun:2010pg,Rijken:2010zza,Yu:2011ta,Zhou:2011sp,Ye:2012gu,Wang:2012wa,
Sun:2013qca,He:2013ttg,Sun:2014wea,Pang:2014laa,Wang:2014sea,Chen:2015iqa}.


This paper is organized as follows. After Introduction,
in {Sec. \ref{sec2}}, we explain the modified Godfrey-Isgur model and the QPC model.
In {Sec. \ref{sec3}}, we adopt the modified Godfrey-Isgur model by including the screening effect
to study the mass spectra of the $K_2^*$ family. We further obtain the structure information of the observed $K_2^*$ via making a comparison between theoretical and experimental results. And we present the detailed study of the OZI-allowed two-body strong decays of the discussed kaons.
The paper ends with a conclusion.

\section{phenomenological analyze of $K_2^*$ mesons}\label{sec2}

In this work, the modified GI quark model is utilized to calculate the mass spectrum and wave functions of the {\color{black}$K_2^*$} meson family. We also investigate the two body strong decay of $K_2^*$ meson family with QPC model.
In the following, these models will be illustrated in details.
\subsection{The brief review of MGI and QPC models}
\subsubsection{The MGI model}
 {\color{black}Godfrey and Isgur {\color{black}raised} the GI model for describing relativistic meson spectra with great success, exactly in low-lying mesons \cite{Godfrey:1985xj}.} As for the excited states, the screening potential must be taken into account for coupled-channel effect \cite{Eef2003ObservedDs,2009Light,2008Contribution}.

The interaction between quark and antiquark is depicted by the Hamiltonian of potential model including kinetic energy pieces and effective potential piece,
\begin{equation}\label{Hamtn}
  \tilde{H}=\sqrt{m_1^2+\mathbf{p}^2}+\sqrt{m_2^2+\mathbf{p}^2}+\tilde{V}_{\mathrm{eff}}(\mathbf{p,r}),
\end{equation}
where $m_1$ and $m_2$ denote the mass of quark and antiquark respectively, and effective potential $\tilde{V}_{\mathrm{eff}}$ contains two ingredients, a short-range $\gamma^{\mu}\otimes\gamma_{\mu}$ one-gluon-exchange interaction and a $1\otimes1$ linear confinement interaction. The meaning of tilde will be explained later.

In the nonrelativistic limit, effective potential has familiar format \cite{Godfrey:1985xj,Lucha:1991vn}
\begin{eqnarray}
V_{\mathrm{eff}}(r)=H^{\mathrm{conf}}+H^{\mathrm{hyp}}+H^{\mathrm{so}}\label{1},
\end{eqnarray}
with
\begin{align}
 H^{\mathrm{conf}}&=\Big[-\frac{3}{4}(c+br)+\frac{\alpha_s(r)}{r}\Big](\bm{F}_1\cdot\bm{F}_2)\nonumber\\ &=S(r)+G(r)\label{3}\\
H^{\mathrm{hyp}}&=-\frac{\alpha_s(r)}{m_{1}m_{2}}\Bigg[\frac{8\pi}{3}\bm{S}_1\cdot\bm{S}_2\delta^3 (\bm r) +\frac{1}{r^3}\Big(\frac{3\bm{S}_1\cdot\bm r \bm{S}_2\cdot\bm r}{r^2} \nonumber  \\ \label{3.1}
&\quad -\bm{S}_1\cdot\bm{S}_2\Big)\Bigg] (\bm{F}_1\cdot\bm{F}_2),  \\
H^{\mathrm{so}}=&H^{\mathrm{so(cm)}}+H^{\mathrm{so(tp)}},  \label{3.2}
\end{align}
where $H^{\mathrm{conf}}$ includes the spin-independent linear confinement piece $S(r)$ and Coulomb-like potential from one-gluon-exchange $G(r)$. $H^{\mathrm{hyp}}$ denotes the color-hyperfine interaction consisting tensor and contact terms. 
 $H^{\mathrm{SO}}$ is the spin-orbit interaction with
\begin{eqnarray}
H^{\mathrm{so(cm)}}=\frac{-\alpha_s(r)}{r^3}\left(\frac{1}{m_{1}}+\frac{1}{m_{2}}\right)\left(\frac{\bm{S}_1}{m_{1}}+\frac{\bm{S}_2}{m_{2}}\right)
\cdot
\bm{L}(\bm{F}_1\cdot\bm{F}_2),
\end{eqnarray}
which is caused by one-gluon-exchange and
\begin{eqnarray}
H^{\mathrm{so(tp)}}=-\frac{1}{2r}\frac{\partial H^{\mathrm{conf}}}{\partial
r}\Bigg(\frac{\bm{S}_1}{m^2_{1}}+\frac{\bm{S}_2}{m^2_{2}}\Bigg)\cdot \bm{L},
\end{eqnarray}
which is the Thomas precession term.
\par
    For above formulas, $\bm{S}_1/\bm{S}_2$ indicates the spin of quark/antiquark and $\bm{L}$ {\color{black}is} the orbital momentum between them. $\bm{F}$ is relevant to the Gell-Mann matrix, i.e., $\bm{F}_1=\bm{\lambda}_1/2$ and $\bm{F}_2=-\bm{\lambda}^*_2/2$, and for a meson, $\langle\bm{F}_1\cdot\bm{F}_2 \rangle=-4/3$.
\par
     Now relativistic effects of distinguish influence must be considered especially in meson system, which is embedded in two ways. Firstly, based on the nonlocal interactions and new $\mathbf{r}$ dependence, a smearing function is introduced for a meson $q\bar{q}$
\begin{equation}
\rho \left(\mathbf{r}-\mathbf{r'}\right)=\frac{\sigma^3}{\pi ^{3/2}}e^{-\sigma^2\left(\mathbf{r}-\mathbf{r'}\right)^2},
\end{equation}
which is applied to $S(r)$ and $G(r)$ to obtain smeared potentials $\tilde{S}(r)$ and $\tilde{G}(r)$ by
\begin{equation}\label{smear}
\tilde{f}(r)=\int d^3r'\rho(\mathbf{r}-\mathbf{r'})f(r'),
\end{equation}
with
\begin{eqnarray}
   \sigma_{12}^2=\sigma_0^2\Bigg[\frac{1}{2}+\frac{1}{2}\left(\frac{4m_1m_2}{(m_1+m_2)^2}\right)^4\Bigg]+
  s^2\left(\frac{2m_1m_2}{m_1+m_2}\right)^2,
\end{eqnarray}

Secondly, owning to relativistic effects, a general potential should rely on the mass-of-center of interacting quarks. Momentum-dependent factors which will be unity in the nonrelativistic limit are applied as
\begin{equation}
\tilde{G}(r)\to \left(1+\frac{p^2}{E_1E_2}\right)^{1/2}\tilde{G}(r)\left(1 +\frac{p^2}{E_1E_2}\right)^{1/2},
\end{equation}
and
\begin{equation}
  \frac{\tilde{V}_i(r)}{m_1m_2}\to \left(\frac{m_1m_2}{E_1E_2}\right)^{1/2+\epsilon_i} \frac{\tilde{V}_i(r)}{m_1 m_2} \left( \frac{m_1 m_2}{E_1 E_2}\right)^{1/2+\epsilon_i},
\end{equation}
where $\tilde{V}_i(r)$ delegates the contact, tensor, vector spin-orbit and scalar spin-orbit terms, and $\epsilon_i$ is the relevant modification parameters.

 The screen effect is considered to be very important for the higher excitation states by the authors of Ref. \cite{Song:2015nia}. It could be introduced by the transformation
 $br+c\rightarrow \frac{b(1-e^{-\mu r})}{\mu}+c$, {\color{black}where $\mu$ is} screened parameter whose particular value is given by Ref. \cite{Pang:2018gcn}. Modified confinement potential also requires similar relativistic correction, which has been mentioned in the GI model.
 Then, we further write
\begin{eqnarray}
\tilde V^{\mathrm{scr}}(r)&=& \int d^3 \bm{r}^\prime
\rho (\bm{r-r^\prime})\frac{b(1-e^{-\mu r'})}{\mu}\nonumber\\
&=& \frac{b}{\mu r}\Bigg[r+e^{\frac{\mu^2}{4 \sigma^2}+\mu r}\frac{\mu+2r\sigma^2}{2\sigma^2}\Bigg(\frac{1}{\sqrt{\pi}}
\int_0^{\frac{\mu+2r\sigma^2}{2\sigma}}e^{-x^2}dx-\frac{1}{2}\Bigg) \nonumber\\
&&-e^{\frac{\mu^2}{4 \sigma^2}-\mu r}\frac{\mu-2r\sigma^2}{2\sigma^2}\Bigg(\frac{1}{\sqrt{\pi}}
\int_0^{\frac{\mu-2r\sigma^2}{2\sigma}}e^{-x^2}dx-\frac{1}{2}\Bigg)\Bigg]
.\label{5}
\end{eqnarray}

The mass spectrum and the wave function of $K_2^*$ mesons can be obtained by solving eigen value and eigen vector of the $\tilde{H}$ in Eq. (\ref{Hamtn}) with the Simple Harmonic Oscillator(SHO) base expanding method.  In configuration and momentum space, SHO wave functions have explicit form respectively
\begin{align}
\Psi_{nLM_L}(\mathbf{r})=R_{nL}(r, \beta)Y_{LM_L}(\Omega_r),\nonumber\\
\Psi_{nLM_L}(\mathbf{p})=R_{nL}(p, \beta)Y_{LM_L}(\Omega_p),
\end{align}
with
\begin{eqnarray}
&R_{nL}(r,\beta)=\beta^{3/2}\sqrt{\frac{2n!}{\Gamma(n+L+3/2)}}(\beta r)^{L}
e^{\frac{-r^2 \beta^2}{2}} \nonumber \\
 &\times L_{n}^{L+1/2}(\beta^2r^2),\\
 &R_{nL}(p,\beta)=\frac{(-1)^n(-i)^L}{ \beta ^{3/2}}e^{-\frac{p^2}{2 \beta ^2}}\sqrt{\frac{2n!}{\Gamma(n+L+3/2)}}{(\frac{p}{\beta})}^{L} \nonumber \\
 &\times L_{n}^{L+1/2}(\frac{p^2}{ \beta ^2}),
\end{eqnarray}
where $Y_{LM_L}(\mathrm{\Omega})$ is spherical harmonic function, and $L_{n-1}^{L+1/2}(x)$ is the associated Laguerre polynomial.

\subsubsection{QPC model}
The QPC model is firstly proposed by Micu \cite{Micu:1968mk}, which is further developed by Orsay group. \cite{LeYaouanc:1972ae,LeYaouanc:1973xz,LeYaouanc:1974mr,LeYaouanc:1977gm,LeYaouanc:1977ux}.
and was widely applied to the OZI-allowed two-body strong decay of hadrons in Refs. \cite{vanBeveren:1979bd,vanBeveren:1982qb,Capstick:1993kb,Page:1995rh,Titov:1995si,Ackleh:1996yt,Blundell:1996as,
Bonnaz:2001aj,Zhou:2004mw,Lu:2006ry,Zhang:2006yj,Luo:2009wu,Sun:2009tg,Liu:2009fe,Sun:2010pg,Rijken:2010zza,Ye:2012gu,
Wang:2012wa,He:2013ttg,Sun:2013qca,Pang:2014laa,Wang:2014sea,Feng:2021igh}.

A decay process $A\to B+C$ can be expressed as
\begin{eqnarray}
\langle BC|\mathcal{T}|A \rangle = \delta ^3(\mathbf{P}_B+\mathbf{P}_C)\mathcal{M}^{{M}_{J_{A}}M_{J_{B}}M_{J_{C}}},
\end{eqnarray}
where $\mathbf{P}_{B(C)}$ is a three-momentum of a meson $B(C)$ in the rest frame of a meson $A$. A superscript $M_{J_{i}}\, (i=A,B,C)$ denotes an orbital
magnetic momentum. The transition operator $\mathcal{T}$ is introduced to describe a quark-antiquark pair creation from vacuum, which has the quantum number
$J^{PC}=0^{++}$, i.e., $\mathcal{T}$ can be written as
\begin{eqnarray}
\mathcal{T}& = &-3\gamma \sum_{m}\langle 1m;1~-m|00\rangle\int d \mathbf{p}_3d\mathbf{p}_4\delta ^3 (\mathbf{p}_3+\mathbf{p}_4) \nonumber \\
 && \times \mathcal{Y}_{1m}\left(\frac{\textbf{p}_3-\mathbf{p}_4}{2}\right)\chi _{1,-m}^{34}\phi _{0}^{34}
\left(\omega_{0}^{34}\right)_{ij}b_{3i}^{\dag}(\mathbf{p}_3)d_{4j}^{\dag}(\mathbf{p}_4).
\end{eqnarray}
It is completely constructed in the form of a visual representation to reflect the creation of a quark-antiquark pair from vacuum, where the quark and antiquark are denoted by indices $3$ and $4$, respectively.
The parameter $\gamma$ depicts the strength of the creation of $q\bar{q}$ from vacuum. $\mathcal{Y}_{\ell m}(\mathbf{p})={|\mathbf{p}|^{\ell}}Y_{\ell
m}(\mathbf{p})$ are the solid harmonics. $\chi$, $\phi$, and $\omega$ denote the spin, flavor, and color wave functions respectively, which can be treated separately.
Subindices $i$ and $j$ denote the color of a $q\bar{q}$ pair.

The decay amplitude can be expressed as another form by the Jacob-Wick formula \cite{Jacob:1959at}
\begin{eqnarray}
\mathcal{M}^{JL}(\mathbf{P})&=&\frac{\sqrt{4\pi(2L+1)}}{2J_A+1}\sum_{M_{J_B}M_{J_C}}\langle L0;JM_{J_A}|J_AM_{J_A}\rangle \nonumber \\
&&\times \langle J_BM_{J_B};J_CM_{J_C}|{J_A}M_{J_A}\rangle \mathcal{M}^{M_{J_{A}}M_{J_B}M_{J_C}},
\end{eqnarray}
then the general decay width will be
\begin{eqnarray}
\Gamma&=&\frac{\pi}{4} \frac{|\mathbf{P}|}{m_A^2}\sum_{J,L}|\mathcal{M}^{JL}(\mathbf{P})|^2,
\end{eqnarray}
where $m_{A}$ is the mass of an initial state $A$.
In our calculation, the spatial wave functions of the mesons are given in Ref. \cite{Pang:2018gcn}. The value of $\gamma$ is 11.6.

\subsection{Mass spectrum analysis}

In Tab. \ref{fitmeason}, we give the mass spectrum of $K_2^*$ mesons
  by using different models. We can see that our previous work Ref. \cite{Pang:2017dlw} give the ground state of $K_2^*$(1P) has a mass of 1432 MeV, which is close to the result of the experimental data \cite{Zyla:2020zbs}.
  \begin{table}[htbp]
\caption{The spectrum of the $K^*_2$ meson {\color{black}family, where} $\text{Exp.}$ represent the  experimental data~\cite{Agashe:2014kda}.
The unit of the mass is MeV. \label{fitmeason}}
\vspace{-16pt}
\[\begin{array}{cccccccc}
\hline
\hline
\text{State}  & \text{Ref.~\cite{Pang:2018gcn}}& \text{Ref.~\cite{Pang:2017dlw}} & \text{GI~\cite{Godfrey:1985xj}} &\text{Ebert~\cite{Ebert:2009ub}}& \text{Exp.} \\
 \hline

  1^3 \text{P}_2 &1450& 1432 & 1409 &1424&  1427 \pm1.5\\
  2^3 \text{P}_2 &1906& 1870 & 1924 & 1896& 1868\pm8^{+40}_{-51} \\
   3^3 \text{P}_2 &2274&  2198 & 2370& --& --\\
  4^3 \text{P}_2 &2570& 2438 & 2756 &-- & -- \\
  1^3 \text{F}_2 &2200& 2092 & 2168 &1.964   & 2073\pm94^{+245}_{-240}\\
  2^3 \text{F}_2 &2415& 2356 & 2565 &-- &-- \\
   3^3 \text{F}_2 &2682& 2552 & 2917 & --& -- \\
\hline
\hline
\end{array}\]
\end{table}For the highly excited state of the P-wave $K_2^*$ mesons, $K_2^*$(3P) and $K_2^*$(4P), they have the mass of 1870 MeV, 2198 MeV and 2438 MeV, respectively, which are smaller than those reported in Ref. \cite{Godfrey:1985xj}. For the F-wave $K_2^*$, $K_2^*$(1F) is predicted with the mass 2092 MeV, which is close to the LHCb data $2073 \pm 94^{+ 245}_{- 240}$ in Ref. \cite{LHCb:2016nsl}.
$K_2^*$(2F) has a mass of 2356 MeV and $K_2^*$(3F) has a mass of 2552 MeV, which are smaller than those reported in Ref. \cite{Godfrey:1985xj}. Additionally, the results of $K_2^*$(1P), $K_2^*$(2P) and $K_2^*$(1F) in Ref. \cite{Ebert:2009ub} are also close to the experiment data. 

\section{Two body strong decay analyze}\label{sec3}

\subsection{{\color{black}The $2^{3}P_2$} and $1^3F_2$ states of $K_2^*$}

\begin{table}[]
\caption{The allowed partial strong decay widths of $2^3P_2$ and $1^3F_2$ state.\label{decay}}
\vspace{-16pt}
\renewcommand{\arraystretch}{1.1}
\[\begin{array}{|c|c|c|}
\hline
\text{Decay channels} & \text{$2^3P_2$} & \text{$1^3F_2$} \\
\hline
 Kb_1 &  {8.15 \sim  26.3} &  {118 \sim  145} \\
 K_1\pi  &  {12.4 \sim  33.8} &  {104 \sim  152} \\
 Ka_1 &  {4.3 \sim  13.8} &  {61.8 \sim  79.8} \\
 K^*\rho  &  {55.8 \sim  77.8} &  {24.1 \sim  131} \\
 
 Kh_1 &  {5.51 \sim  10.8} &  {39.5 \sim  47.5} \\

 K_2{}^*\pi  &  {15.2 \sim  28.3} &  {18.9 \sim  34.4} \\
 Ka_2 &  {0 \sim  25.9} &  {14 \sim  32.4} \\
 Kf_1 &  {0.23 \sim  3.2} &  {15.7 \sim  25.1} \\
 K^*\omega  &  {18.7 \sim  24.9} &  {7.64 \sim  42.3} \\
  {K \rho  } &  {31.6 \sim  33} &  {19.1 \sim  21.5} \\
 K^*(892)\pi  &  {18.1 \sim  21.8} &  {16.8 \sim  20.1} \\

  {K \pi  } &  {0.0117 \sim  1.17} &  {13.9 \sim  19.5} \\

  {K(1460) \pi  } &  {5.15 \sim  21.7} &  {9.26 \sim  16.6} \\

 Kf_2(1270) &  {1.29 \sim  12.4} &  {6.03 \sim  11.2} \\
  {K \omega  } &  {10.5 \sim  11} &  {6.43 \sim  7.19} \\
  {K \pi  (1300)} &  {0\sim  10.5} &  {3.02 \sim  9.17} \\

 K^* {(1410) \pi  } &  {15.2 \sim  46.7} &  {3.37 \sim  5.7} \\
 K^*\eta  &  {3.93 \sim  5.02} &  {3.79 \sim  4.47} \\
  {K \eta  } &  {0.361 \sim  0.801} &  {3.77 \sim  4.4} \\
  {K \eta  '} &  {0.632 \sim  0.818} &  {2.31 \sim  3.27} \\

 K_1^{'}\pi  &  {12.5 \sim  16.9} &  {0.113 \sim  0.498} \\
 K_1\rho  & 0 &  {0 \sim  149} \\
  K_1\eta  & 0 &  {24.3 \sim  37.9} \\
   K\eta _2 & 0 &  {0 \sim  62.8} \\
    K_1\omega  & 0 &  {0 \sim  48.1} \\
     K^*h_1 & 0 &  {0 \sim  34.2} \\
 K^*b_1 & 0 &  {0 \sim  61.4} \\
  K^*a_1 & 0 &  {0 \sim  33.7} \\
   K_2{}^*\eta  & 0 &  {0 \sim  4.99} \\
    K^* { \eta  '} & 0 &  {0.319 \sim  2.45} \\
     K_3{}^* {(1780) \pi  } & 0 &  {0.0372 \sim  7.77} \\
 K^* {(1680) \pi  } & 0 &  {0.0562 \sim  3.73} \\
\hline
\end{array}
\]

\end{table}

When we use the experimental value   $1868 \pm 8^{+ 40}_{- 57}$ MeV and  $M=2073 \pm 94$  MeV(here, we do not take into account the large systematic error part of $K_2^*(2070)$) as input for the mass of $2^3P_2$ and $1^3F_2$ state,   
the QPC model provides us an effective approach to obtain the decay width of the $2^3P_2$ and $1^3F_2$ state, which are $285 \pm 45$ and $855 \pm 225$ MeV respectively. In order to clearly compare our prediction with the resonance parameters of  $K_2^*(1980)$ measured by different experiments collected in PDG \cite{Zyla:2020zbs}, we present the total width and decay branching ratio of $K_2^*(1870)$ and $K_2^*(2070)$ with the variation of the mass. 
\begin{figure}
\includegraphics[width=6cm]{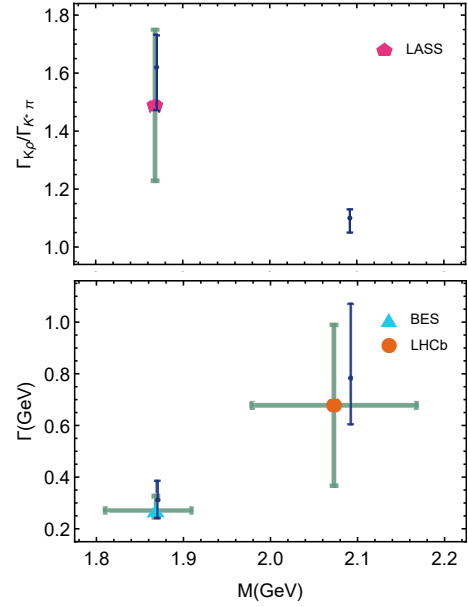}
\caption{Our theoretical results and the resonance parameters of $K_2^*(1980)$ measured by different experiments collected in PDG \cite{Zyla:2020zbs}. Here, the green lines with triangle and disk denote the central value of mass and decay width of $K_2^*(1980)$ measured by BESIII and LHCb, respectively \cite{BESIII:2019apb, LHCb:2016nsl}. The green line with rose pentagon denotes the central value of decay branching ratio of $K_2^*(1980)$ measured by LASS \cite{ASTON1987693}. The purplish blue lines are our calculation results.}
\label{fpw}
\end{figure}
The comparison between our theoretical result of the total width of $K_2^*(1870)$ with experimental data is shown in the under diagram of Fig. \ref{fpw}, notice that our result is very close to the BESIII experimental  data \cite{BESIII:2019apb} marked by the green lines with triangle in Fig. \ref{fpw}. Based on the decay branching ratio of $\frac {\text {K$ \rho $}} { K^*\pi}$, when $K_2^*(1870)${\color{black} was} regarded as a $ 2^3P_2$ state, our result of $\frac{\Gamma _{\text{K$\rho $}}}{\Gamma _{K^*\pi  }}$ = $1.62^{+0.11}_{-0.15}$ well conform  to the experimental value $\frac{\Gamma _{\text{K$\rho $}}}{\Gamma _{K^*\pi  }}$ = $1.49 \pm 0.24\pm 0.09$ \cite{ASTON1987693}. Thus, explaining $K_2^*(1870)$ as a $ 2^3P_2$ state is further tested through the decay branching ratio of $\frac {\text {K$ \rho $}} { K^*\pi}$. In the upper diagram of Fig. \ref{fpw}, when we treat $K_2^*(2070)$ as 1F state, the ratio of $\frac{\Gamma _{\text{K$\rho $}}}{\Gamma _{K^*\pi  }}$ is about $1.05-1.13$. The width has a overlap with the the LHCb data $\Gamma _{ K_2^*(2070)}$ = $678 \pm 311^{+559}_{-1153}$ MeV in  $M = (1979 - 2167)$ MeV range ($M _{ K_2^*(2070)}$ = $2073 \pm 94^{+245}_{-240}$ MeV \cite{LHCb:2016nsl}). 
The total error  $+640$  MeV is comparable with the centre width 678 MeV, and another total error $-1194$  MeV is nearly two times of this centre width value, this is the largest experimental error for experimental data of $K$ meson family in PDG \cite{Zyla:2020zbs}. Notice that the mass and width of $K^*(2^+) 2^3P_2$ state are fitted using the quantum numbers $n^{2S+1}L_J=2^3P_2$ in Ref. \cite{LHCb:2016nsl}. Additionally, we also note that PDG edition 2022 not adopt this LHCb data \cite{Workman:2022ynf}, maybe this “not adopt” caused by the large error. So we suggest the experimenters add the $K_2^*(1^3F_2)$ state in the fit scheme in \cite{LHCb:2016nsl}, which is possible to reduce the experimental error of $K_2^*$.

We have a nice description of $K_2^*(1870)$ on the mass spectrum and decay behaviors under the assignment of $K_2^*$ $(2^3P_2)$.
The assignment of $K_2^*$ $(1^3F_2)$ for  $K_2^*(2070)$ need more experimental support.
We hope our theoretical result can help to establish this $K_2^*(1F)$ state.

The two-body decay information of $2^{3}P_2$ and $1^{3}F_2$ state can be obtained in Tab. \ref{decay}. $Kb_1$, $K_1 \pi$, $Ka_1$ and $K^* \rho$ all made important contribution to $2^{3}P_2$ and $1^{3}F_2$ state. The decay mode $K_1 \rho$, $K_1 \eta$, $K_1 \eta_2$ and $K_1 \omega$ is predicted to be dominant to $1^{3}F_2$ state, but has no contribution to $2^{3}P_2$ state. $Kh_1$, $K_2^* \pi$, $Ka_2$, $Kf_1$, $K^* \omega$, $K \rho$ and $K^* \pi$ have visible contribution to the total width of $2^{3}P_2$ and $1^{3}F_2$ state. $K^* \eta^\prime$, $K_3^*(1780) \pi$ and $K^*(1680) \pi$ have very small widths in the final states of $2^{3}P_2$ and $1^{3}F_2$ state. SPEC found an indication of a decay channel of $K_2^* (1980)$: $Kf_2(1270)$ in 2003 \cite{Tikhomirov:2003gg}. The PDG considers only $K \rho$, $K^* \pi$, $Kf_2(1270)$, $K^* \phi$ modes for “$K_2^*(1980)(2P)$” state, and $K_2^*(1980)$ → $K \eta$ are observed for the first time by Y. Q. Chen $et~al.$ in the $D_0$ → $K^- \pi^+$ decays \cite{2020Dalitz}. The width of $Kf_2(1270)$ channel is predicted to be $1.29 - 12.4$ and $6.03-11.2$ MeV for $2^{3}P_2$ and $1^{3}F_2$ state, respectively. The predicted ordering of two widths $K \rho$ $>$ $K^* \pi$ is in agreement with experiment \cite{ASTON1987693}, and the predicted and observed decay branching ratio are roughly consistent. 

The largest channels of $1^{3}F_2$ state are predicted to be $K_1 \rho$, $Kb_1$, $K^* \rho$, $K_1 \pi$ and $Ka_1$, with branching fractions of $1\%$, $7\%$, $3\%$, $7\%$ and $4\%$ respectively when taking  $M = 2070$ MeV. Two of them are larger than $5\%$, these being $Kb_1$ and $K_1 \pi$. Note that some decay channels have strong dependence with the change of  the mass, like $K^*(1680) \pi$, $K^* \eta^\prime$, $K(1460) \eta$, $K \eta(1295)$, $Kf_1(1420)$ and $K^*(1410) \eta$. Observation of these channels liking $K \rho$, $K^* \pi$ and $K\pi$ may provide useful information about the nature of $K_2^*(1980)$ meson.

\subsection{The predicted $K_2^*(3P)$ and $K_2^*(4P)$ states}

\begin{figure*}
\center
\includegraphics[width=15.2cm]{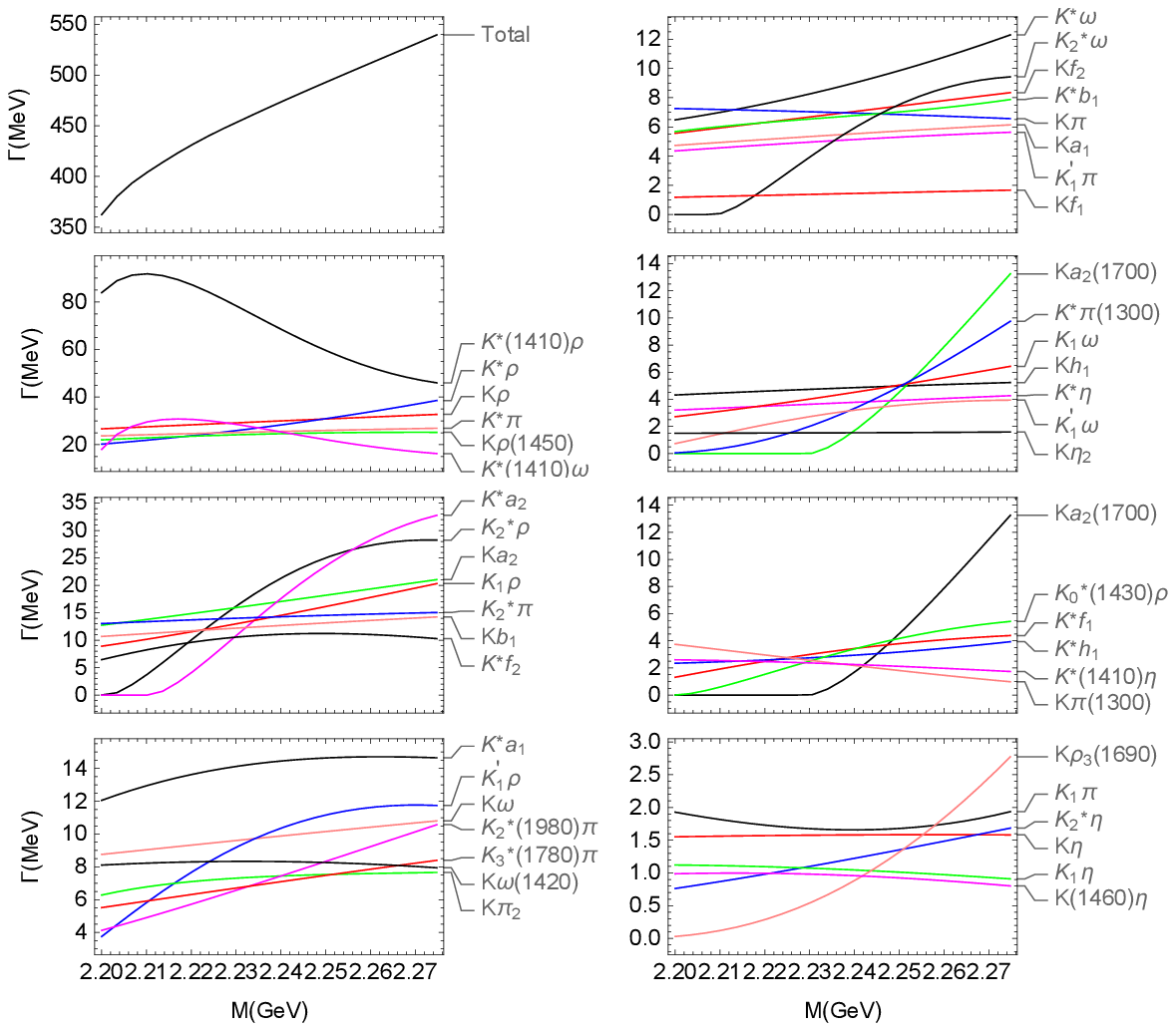}
\caption{The M dependence of the calculated decay widths of $3^{3}P_2$ state.}
\label{3P}
\end{figure*}

When further discussing the decay {\color{black}behavior} of $ 3^3P_2$ {\color{black}state of $K_2^*$ meson family,} we may estimate the total decay width of $K_2^*(3P)$ to be {\color{black}(360 - 540)} MeV and the mass to be $(2200 - 2276)$ MeV. The predicted main decay channels of $K_2^*(3P)$ include $K^*(1410) \rho$, $K^* \rho$, $K \rho$,$K^* \pi$, $K \rho(1450)$,  $K^*(1410) \rho$ channel has regnant position. The channel of $K_1 \pi$, $K_2^* \eta$, $K \eta$, $K_1 \eta$ and $K(1460) \eta$ have very small contribution to the total decay, they {\color{black}are} not sensitive to the change of the mass.  {\color{black}More details can be found in Fig. \ref{3P}.}
\begin{figure*}
\center
\includegraphics[width=15.2cm]{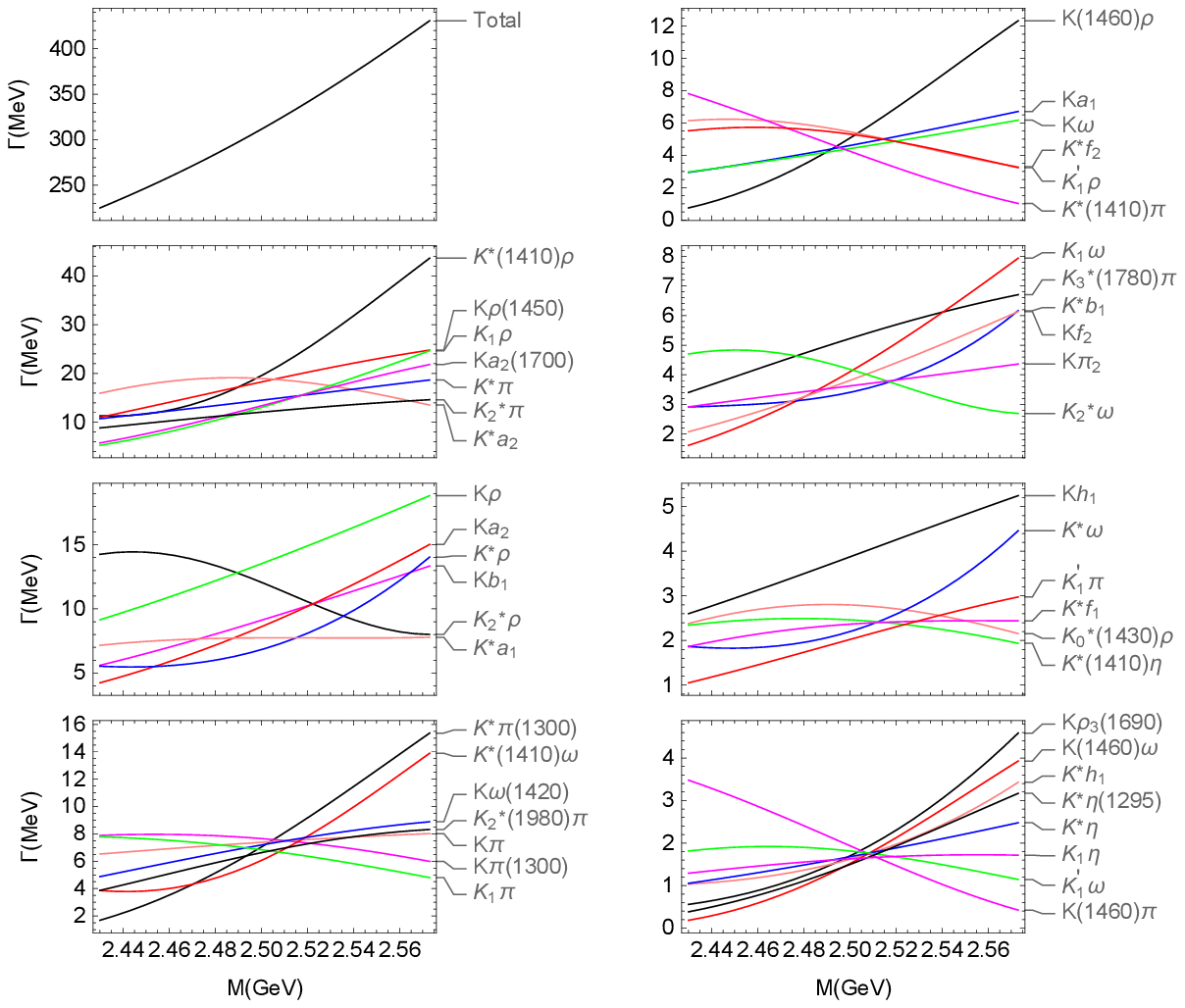}
\caption{The M dependence of the calculated decay widths of $4^{3}P_2$ state.}
\label{4P}
\end{figure*}

The calculated total decay width of $K_2^*(4P)$ is $(225 - 430)$ MeV when taking {\color{black}$M = (2436 - 2566)$ MeV}. It is evident from Fig. \ref{4P} that searches for these decay modes, we find that only a few channels have branching fractions larger than a few percent. $K^*(1410) \rho$, $K \rho(1450)$, $K_1 \rho$, $Ka_2(1700)$, $K^* \pi$,$K_2^* \pi$ and $K^*a_2$ are the main decay modes of $K_2^*(4P)$, which have the branching ratios of $0.04 - 0.10$, $0.04 - 0.06$, $0.03 - 0.06$, $0.03 - 0.05$, 0.04, $0.03 - 0.04$, and $0.03 - 0.07$, respectively.

\subsection{The predicted $K_2^*(2F)$ and $K_2^*(3F)$ states}

\begin{figure*}
\center
\includegraphics[width=15.2cm]{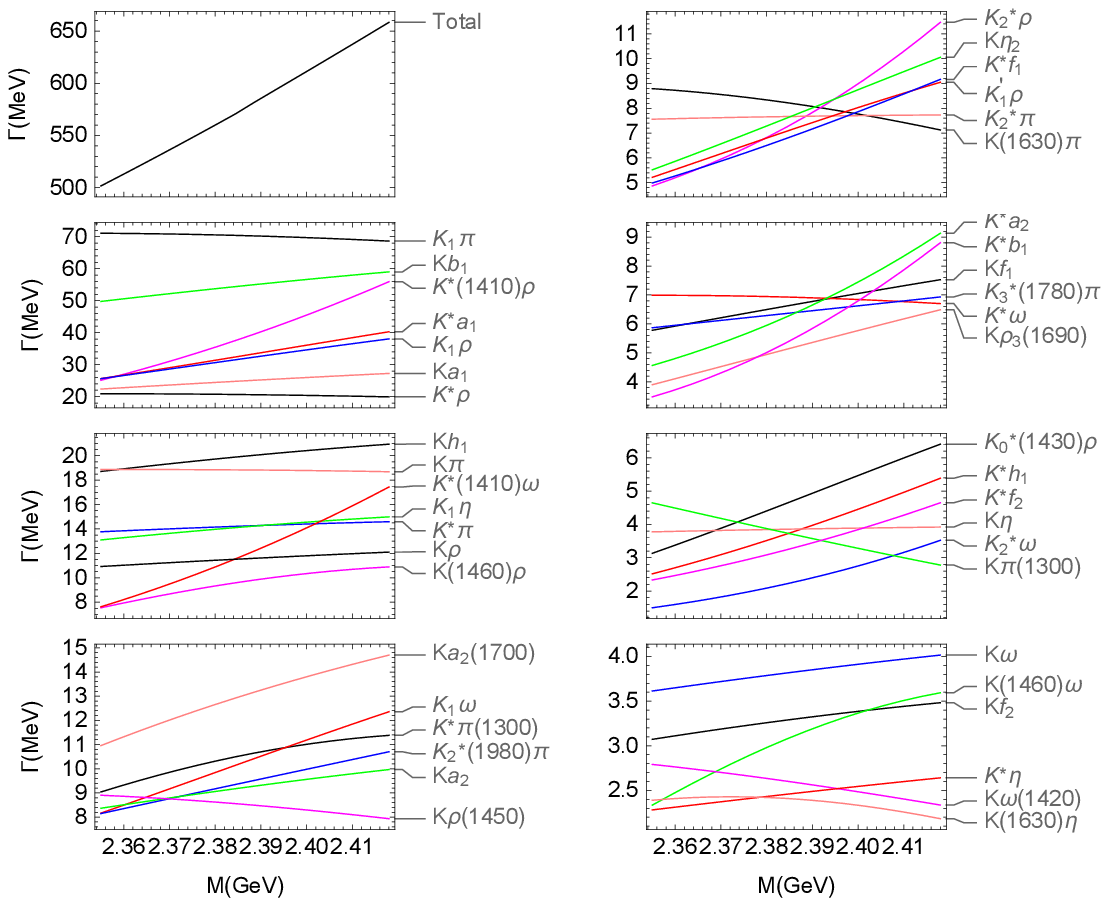}
\caption{The $M$ dependence of the calculated decay widths of $2^{3}F_2$ state.}
\label{2F}
\end{figure*}

 The predicted mass of the {\color{black}$2^3F_2$ state in $K_2^*$ meson family is of $2415$ MeV}. Our result (Fig. \ref{2F})
shows that when we take its mass  2355 MeV  to 2565 MeV, the total width is $\Gamma_{ K_2^*(2^3F_2)}$ = $580 \pm 80$ MeV. The largest decay channel $K_1 \pi$ do not change significantly with the mass in $M = (2355 - 2565)$ MeV range and its branch ratio is about 0.12. The important decay channels are $Kb_1$, $K^*(1410) \rho$, $K^*a_1$, $K_1 \rho$, $Ka_1$, $K^* \rho$. In addition,  $Kh_1$,  $K \pi$, $K^*(1410) \omega$,  $K_1 \eta$, $K^* \pi$, $K \rho$ and $K(1460) \rho$ also have certain contribution. $K \omega$, $K(1460) \omega$, $Kf_2$, $K^* \eta$, $K \omega(1420)$, $K(1630) \eta$ make small contribution.  We hope that the predicted behavior of $K_2^*(2F)$ is {\color{black}helpful} to the experimental search for $K_2^*(2F)$ {\color{black}state}.

\begin{figure*}
\center
\includegraphics[width=14.5cm]{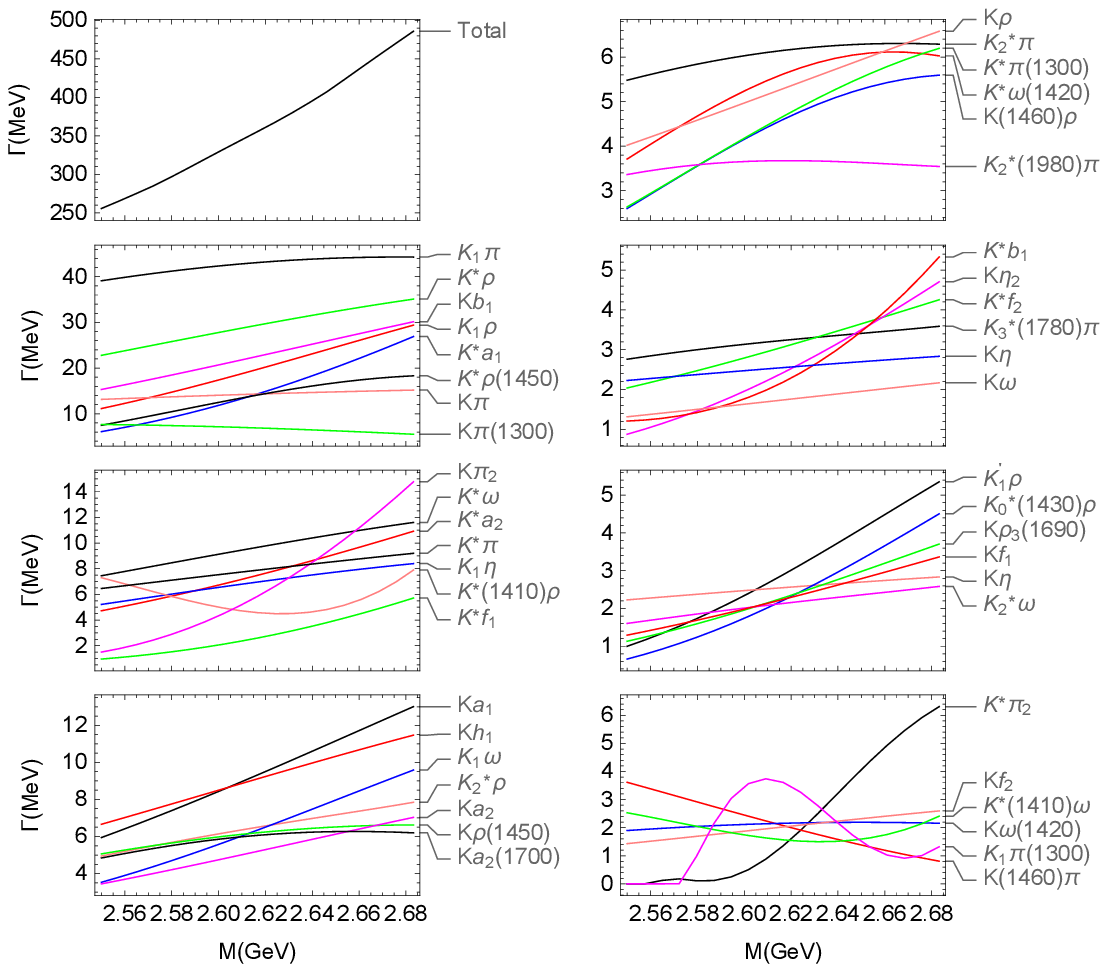}
\caption{The $M$ dependence of the calculated decay widths of $3^{3}F_2$ state.}
\label{3F}
\end{figure*}	

As we can see from Fig. \ref{3F}, comparing to the predictive $K_2^*(1F)$ state, the predictive $K_2^*(2F)$ state and $K_2^*(3F)$ state have more decay channels. The  obtained mass is $M_{3^3F_2}$ = $2624 \pm 58$ MeV. The corresponding total decay width is about $\Gamma_{3^3F_2}$
=$370 \pm 120$ MeV. Note that the important decays are again distributed over several modes, the larger decay modes are $K_1 \pi$, $K^* \rho$,$Kb_1$, $K_1 \rho$,  $K^*a_1$,  $K^* \rho(1450)$,  $K \pi$, and $K \pi(1300)$. $Kf_2$, $K^*(1410) \omega$, $K \omega(1420)$, $K_1\pi(1300)$ and $K(1460) \pi$ contribute very little to the {\color{black}total decay width of $K_2^*(3F)$.}

\section{Conclusion}\label{sec4}

The observed $K_2^*(1870)$ and $K_2^*(2070)$ are firstly explained as $2^3P_2$ and $1^3F_2$ state respectively. By analyzing the mass spectra of the P-wave and the F-wave $K_2^*$ meson family and calculating the two body strong decay of this two states, we find that our predicted results of $K_2^*(1870)$ are consistent well with the existed experimental findings. Our results about $K_2^*(2070)$ have a large overlap with existed experimental findings. Our theoretical results show that, $K_2^*(1870)$ can be regard as a $2^3P_2$ state through the comparison with the experimental data. $K_2^*(2070)$ is likely to be a  $1^3F_2$ state. $1^3F_2$  state may has a relatively large width of $855 \pm 225$ MeV, and the ratio of $\frac{\Gamma _{\text{K$\rho $}}}{\Gamma _{ K^*\pi }}$ is $1.05-1.13$. Just because of our explanations to $K_2^*(1870)$ and $K_2^*(2070)$, the spectroscopy of P-wave and the F-wave $K_2^*$ mesons become abundant. Additionally, we predict the decay behaviors of the $K_2^*(2P)$ and $K_2^*(1F)$ state and the decay width of channels like $K \rho$, $K^* \pi$ and $Kf_2(1270)$ are calculated.  These findings are expected to be revealed in future experiment. Besides the $2^{3}P_2$ and $1^3F_2$ strange mesons, the decay behaviors of the other higher excited  $K_2^*$ mesons have been also predicted in the present work. The masses and widths of these predicted states provide
some basic information that may help for searching for these
strange mesons in future experiments.

Note added: We hope that the resonance parameter (total width) of $K_2^*(2070)$ could be fitted again by experimental group considering the  quantum numbers $n^{2S+1}L_J=1^3F_2$ of $K_2^*$, which will provide the powerful criterion to test the information for confirming $K_2^*(2070)$ state further.

\section{Acknowledgments}
This  work is supported in part by National Natural Science Foundation of China under the
Grant No. 11965016, Qinghai Science and Technology Plan, No. 2020-ZJ-728. C. Q. thanks Jun-zhang Wang for the useful discussions.
\vfil
\bibliographystyle{apsrev4-1}
\bibliography{hep}
\end{document}